\documentstyle[12pt]{article}
\def\beq{\begin{equation}}
\def\eeq{\end{equation}}
\def\bea{\begin{eqnarray}}
\def\eea{\end{eqnarray}}

\def\ba{\begin{array}}
\def\ea{\end{array}}
\def\bce{\begin{center}}
\def\ece{\end{center}}

\begin{document}
\begin{titlepage}
\rightline{APCTP-97-08}
\def\today{\ifcase\month\or
        January\or February\or March\or April\or May\or June\or
        July\or August\or September\or October\or November\or December\fi,
  \number\year}
\rightline{hep-th/9705004}
\vskip 1cm
\centerline{\Large \bf Geometry, D-Branes and  $N=1$ Duality in 
Four Dimensions II}
\vskip 2cm
\centerline{\sc Changhyun Ahn \footnote{chahn@apctp.kaist.ac.kr}}
\centerline {{\it Asia Pacific Center for Theoretical Physics (APCTP)}}
\centerline{{\it 207-43 Cheongryangri-dong, Dongdaemun-gu}}
\centerline {{\it Seoul 130-012, Korea}}
\vskip 2cm
\centerline{\sc Abstract}
\vskip 0.2in
We study $N=1$ dualities in four dimensional supersymmetric gauge theories
in terms of  wrapping D 6-branes around 3-cycles of
Calabi-Yau threefolds in type IIA string theory. 
We generalize the recent work of geometrical realization 
for the models which have the superpotential 
corresponding to an $A_k$ type singularity, to various models
presented by Brodie and Strassler, consisting of 
$D_{k+2}$ superpotential of the form $W=\mbox{Tr} X^{k+1} + \mbox{Tr} XY^2$.
We discuss a large number of representations for the field $Y$, but with $X$
always in the adjoint (symmetric) [antisymmetric] representation for $SU (SO) [Sp]$
gauge groups.
\vskip 0.8in
\leftline{May, 1997}
\end{titlepage}
\newpage

\section{ Introduction and Geometrical Setup}
\setcounter{equation}{0}

String theory interprets many nontrivial aspects of four dimensional
supersymmetric field theory by exploiting T-duality on the
local model for the compactification manifold.
There are two approaches to describe this local model. 

Approach I) is to consider the local description as purely geometric
structure of compactification manifold together with D-branes wrapping
around cycles \cite{KV1,BJPSV,VZ}.
The compactification of F-theory on elliptic 
Calabi-Yau(CY) fourfolds from 12 dimensions leads to
$N=1$ supersymmetric field theories in four dimensions.
It has been studied in \cite{KV1},
for the case of pure $SU(N_c)$ Yang-Mills gauge theory,
that the gauge symmetry can be obtained in terms of the structure of
the D 7-brane worldvolume.
By adding some numbers of D 3-branes and bringing them
near the complex 2-dimensional surface (which is the 
compact part of D 7-brane worldvolume),   
the local string model gives rise to matter hypermultiplets in the 
fundamental representation with
pure $SU(N_c)$ Yang-Mills theory \cite{BJPSV}. 
Moreover, Seiberg's duality \cite{Seiberg} 
for the $N=1$
supersymmetric field theory can be mapped to T-duality exchanging the
D 3-brane charge and D 7-brane charge. 
For the extension of $SO(N_c)$ and $Sp(N_c)$ 
gauge theories \cite{IS,IP} coupled
to matter, the local string models are type IIB orientifolds, for which 
T-duality symmetry applies, with D 7-branes on a curved orientifold
7-plane \cite{VZ}.

Approach II) which is T-dual to the approach I is 
to interpret $N=1$ duality for $SU(N_c)$ gauge theory
as D-brane description together with
NS 5-branes in a flat geometry \cite{EGK} according to the approach of \cite{HW}. 
Extension of this to the case of $SO(N_c)$ and $Sp(N_c)$
gauge theories with flavors was presented in \cite{EJS} by
adding an orientifold 4-plane.
The generalization to the construction of product gauge groups $SU(N_c) \times
SU(N'_c)$ with matter fields is given in \cite{BH,BSTY} by suspending two sets of
D 4-branes between three NS 5-branes.
See also the recent paper \cite{Tatar} dealing with $ SO(N_c) \times Sp(N'_c)$ 
product gauge group.

Ooguri and Vafa have considered in \cite{OV} that
$N=1$ duality can be embedded into 
type IIA string theory with D 6-branes, partially
wrapped around three cycles of CY threefold, filling four
dimensional spacetime.
They discussed in the spirit of \cite{BJPSV,VZ}
what happens to the wrapped cycles and studied
the relevant field theory results when a transition in  the moduli of CY 
threefolds occurs. Furthermore, they reinterpreted the configuration of
D-branes in the presence of NS 5-branes \cite{EGK} 
as purely classical geometrical
realization.

It is natural to ask how a number of known field theory dualities
which contain {\it additional} field contents
arise from two different approaches we have described so far. 
Recently, it was obtained in \cite{AO} according to the approach I that 
one can  generalize the approach of \cite{OV} to various models,
consisting of one or two 2-index tensors 
and some fields in the defining
representation (fundamental representation for $SU(N_c)$ and $Sp(N_c)$,
vector representation for $SO(N_c)$), presented earlier by many 
authors \cite{Kutasov,Intril,LS,ILS} and study its geometric realizations
by wrapping D 6-branes about 3-cycles of CY threefolds. 
On the other hand, very recently by introducing a multiple of coincident NS 5-branes
and that of NS' 5-branes, it was argued \cite{EGKRS} that $SU(N_c)$ gauge
theory with one or two adjoints superfields in addition to the fundamentals can be
obtained. Moreover, $SO(N_c)$ gauge group with an adjoint field and $N_f$ vectors
was described in terms of a multiple of coincident NS 5-branes and a single NS'
5-brane with orientifold 4-plane( similarly $Sp(N_c)$ with a traceless antisymmetric
tensor and $N_f$ flavors by adding orientifold 6-plane).
See also the relevant paper \cite{AH} analyzing the brane configurations
associated with field theories in various dimensions.

In this paper, we apply the approach I to the various models
presented by Brodie and Strassler \cite{BS}, consisting of 
$D_{k+2}$ superpotential with $SU, SO$ and $Sp$ gauge groups along the line of
\cite{AO}.
We discuss a large number of representations for the field $Y$, but with $X$
always in the adjoint (symmetric) [antisymmetric] representation for $SU (SO) [Sp]$
gauge groups. The superpotentials are given by 
$W=\mbox{Tr} X^{k+1} + \mbox{Tr} XY^2$ in the three cases while those in other
two cases have different forms that will appear later.

We will  review the main geometrical setup of \cite{OV,AO} in the remaining part of
this section.
Let us start with the compactification of type IIB string theory
on the CY threefold leading to $N=2$ supersymmetric field theories in 4 
dimensions. Suppose various D 3-branes wrapping around
a set of three cycles of CY threefold. It is 
known \cite{CDFV} that whenever
the integration of the holomorphic 3-form on the CY threefold 
around three cycles  
takes the form of parallel vectors in the complex plane, such a D 3-brane configuration
allows us to have a BPS state.
Then after we do T-dualizing the 3-spatial directions of three torus
$T^3$ we obtain type IIA string theory with D 6-branes, partially
wrapping around three cycles of CY threefold,
filling 4 dimensional spacetime. 
We end up with 
$N=1$ supersymmetric field theories in 4 dimensions. 

The local model of CY threefold can be described by \cite{BSV,OV}
five complex coordinates $x, y, x', y'$ and $z$ satisfying the following
equations:
$$
x^2+y^2=\prod_i (z-a_i), \;\;\;\; x'^2+y'^2=\prod_j (z-b_j)
$$ 
where each of $C^*$'s is embedded in $(x, y)$-space and $(x', y')$-space
respectively over a generic point $z$. 
This describes a family of a product of
two copies of one-sheeted hyperboloids in $(x, y)$-space 
and $(x', y')$-space respectively parameterized
by the $z$-coordinates. 
For a fixed $z$ away from $a_i$ and $b_j$ there
exist nontrivial $S^1$'s in each of $C^*$'s corresponding to the waist
of the hyperboloids. Notice that
when $z=a_i$ or $z=b_j$ the corresponding circles vanish as the waists shrink.
Then we regard 3 cycles as the product of $S^1 \times S^1$ cycles over
each point on the $z$-plane, with the segments in the $z$-plane ending
on $a_i$ or $b_j$. When we go between two $a_i$'s ($b_j$'s) 
without passing through
$b_j$ ($a_i$) the 3 cycles sweep out $S^2 \times S^1$.
On the other hand, when we go between $a_i$ and $b_j$ the 3 cycle becomes
$S^3$. We will denote the 3-cycle lying over between $a_i$ and $b_j$
by $[a_i, b_j]$ and also denote other cycles in a similar fashion.

In order to study  $SO(N_c)$  and $Sp(N_c)$ gauge theories, as done in\cite{OV,AO}
consider the local model of the CY threefold given by
$$
x^2+y^2= -\prod_i (z-a_i)(z-a'), \;\;\;\;  x'^2+y'^2=-z
$$
where $a_i$'s and $a'$ are real numbers with $a_1< a_2 <\cdots <a_k <0 < a'$.
It is easily checked that the $S^2 \times S^1$ associated with $[a_{i-1}, a_{i}]$ 
for $i< k$ is realized either by real values for $x, y, x', y', z$ or
by purely imaginary values for $x, y, x', y'$ but real values for $z$. 
Also note  that the $S^3$ associated
with $[a_k, 0]$ is realized by taking the real
values of $x, y, x', y'$ and  $z$ while the $S^3$ associated
with $[0, a']$ is realized
by taking the imaginary values of $x, y, x'$ and $y'$  and real
values for $z$.
In next section we will describe our main results by exploiting this geometrical setup.
We will start with by writing down 
the configurations of ordered points in the real axis of $z$-plane to various models
we are concerned with.

\section{Geometrical Realization of $N=1$ Duality}
\setcounter{equation}{0}

Let us consider particular $N=1$ supersymmetric field theories
and see how their $N=1$ dualities arise from the configurations of ordered
points after the  transition in  moduli
space of CY threefolds.
  
{\bf 1) $SU(N_c)$ with two adjoint fields and $N_f$ fundamental 
flavors \cite{Brodie,BS}:}

We study supersymmetric Yang-Mills theory with gauge group $SU(N_c)$ coupled to
two chiral matter superfields $X$ and $Y$ which transform 
under the adjoint representation of $SU(N_c)$,
$N_f$ fundamental chiral multiplets $Q^i$ and $N_f$ antifundamental
multiplets $\widetilde{{Q}_{\widetilde{i}}}$ where $i, 
\widetilde{i}=1, \cdots, N_f$. The superpotential is
$\mbox{Tr} X^{k+1}+\mbox{Tr} XY^2$ where $k$ is odd. 
We will discuss every steps for this case in detail
while the other cases will be explained very concisely.   
Let us consider the configuration of points ordered as
$$\left( a_1, b_1, b_2, b_3, b'_1, b'_2, \cdots, b'_{\frac{3(k-1)}{2}},
 b''_1, b''_2, \cdots, b''_{\frac{3(k-1)}{2}}, a_2 \right)$$
where it is understood that when $k=1$ the only points of $b_1, b_2$ and $b_3$
appear between $a_1$ and $a_2$.
Suppose that
we wrap $N_i$ D-branes around the
cycle $[a_1, b_i]$,  
$N'_j$ D-branes around the
cycle $[a_1, b'_j]$ and $N''_l$ D-branes around 
$[a_1, b''_l]$ for each $i, j$ and $l$
such that $\sum_{i=1}^{3} N_i+\sum_{j=1}^{\frac{3(k-1)}{2}} (N'_j+ N''_j)=N_c$. 
Each  of them has $N_f$ D-branes around the cycles $[b_i, a_2], [b'_j, a_2]$ and
$[b''_l, a_2] $ for $i, j$ and $l$ respectively. 

We now proceed the case of $k=1$ for simplicity and discuss how the $N=1$ duality is
realized geometrically by D-brane picture from our above proposed configuration.
There are two points $a_1$ and $a_2$ along the real
part of $z$-plane where the first $C^*$ degenerates and three
points $b_1, b_2$ and $b_3$ along the real axis between $a_1$ and $a_2$ where
the second $C^*$ degenerates. Now we have five ordered special
points $(a_1, b_1, b_2, b_3, a_2)$ along the real axis. 
Then the three cycle $[a_1, b_1]$ lying between $a_1$ and $b_1$ is $S^3$ 
and the three cycle $[b_1, b_2]$ lying between $b_1$ and $b_2$ is
 $S^1 \times S^2$. Thus the three cycle $[a_1, b_2]$ is a bouquet of $S^3$
and $S^1\times S^2$ joined together at $z=b_1$.
 We wrap $N_i$ D-branes around the
three cycle $[a_1, b_i]$ for each $i=1, 2, 3$
such that $\sum_{i=1}^3 N_i=N_c$ (because in the limit of
$b_1 \rightarrow b_3$ and $b_2 \rightarrow b_3$ 
this system should be consistent with 
$N_c$ D-branes around the cycle $[a_1, b_3]$) 
and $N_f$ D-branes around the three  cycle $[b_i, a_2]$ for each $i$
where we assume that $3 N_f \geq N_c$. 

Now we would like to move to other point in
the moduli of CY threefolds and end up with
the configuration in which the degeneration points are along the real
$z$-axis except that the orders are changed from $(a_1, b_1, b_2, b_3, a_2)$
to $(b_1, b_2, b_3, a_1, a_2)$. As done in \cite{OV,AO}, first of all we push the point $b_1$
up along the imaginary direction since we have the freedom to
turn on a Fayet-Iliopoulos(FI) D term. Then $(N_1+N_2+N_3)$ of D-branes connect
directly between $(a_1, b_2)$ and $(N_2+N_f+N_3-N_1-N_2-N_3)$ of D-branes go
between $(b_1, b_2)$. We continue to move $b_1$ along the negative
real axis, pass the $x$-coordinate of $a_1$ and push down it to the real
axis. At this moment, the $(N_f-N_1)$ D-branes which were between 
$(b_1, b_2)$ decompose to $(N_f-N_1)$ D-branes between $(b_1, a_1)$ and
$(N_f-N_1)$ D-branes between $(a_1, b_2)$ which amounts to the decomposition
of the three cycle $[b_1, b_2]$ into a bouquet of
two 3-cycles of $S^3$. 
The $(N_1+N_2+N_3)$ D-branes which were
going between $(a_1, b_2)$ will recombine with the newly generated
$(N_f-N_1)$ D-branes
ending up with the total of $(N_f+N_2+N_3)$ D-branes along $[a_1, b_2]$ cycle.
Similarly we do push the point $b_2$ in turn and move it between
$b_1$ and $a_1$.  
Then $(N_f+N_2+N_3)$ of D-branes connect
directly between $(a_1, b_3)$ and $(2 N_f+N_3-N_f- N_2-N_3)$ of D-branes go
between $(b_2, b_3)$. 
We can see that the $(N_f-N_2)$ D-branes which were between 
$(b_2, b_3)$ decompose to $(N_f-N_2)$ D-branes between $(b_2, a_1)$ and
$(N_f-N_2)$ D-branes between $(a_1, b_3)$.
The $(N_f+N_2+N_3)$ D-branes which were
going between $(a_1, b_3)$ do recombine with the new $(N_f-N_2)$ D-branes
ending up with the total of $(2 N_f+N_3)$ D-branes along $[a_1, b_3]$ cycle.
Of course, there are $(2 N_f-N_1-N_2)$ D-branes between $(b_2, a_1)$ due to the
two contributions from  $(N_f-N_1)$ D-branes between $(b_1, a_1)$ ( which can be
decomposed into  again $(N_f-N_1)$ D-branes between $(b_1, b_2)$ and those between 
$(b_2, a_1)$ ) and
$(N_f-N_2)$ D-branes between $(b_2, a_1)$.
Finally, we push the point $b_3$ and move it between
$b_2$ and $a_1$.  
Then $(2N_f+N_3)$ of D-branes connect
directly between $(a_1, a_2)$ and $(3 N_f-N_3-2 N_f)$ of D-branes go
between $(b_3, a_2)$. 
It is easy to see that the $(N_f-N_3)$ D-branes which were between 
$(b_3, a_2)$ decompose to $(N_f-N_3)$ D-branes between $(b_3, a_1)$ and
$(N_f-N_3)$ D-branes between $(a_1, a_2)$.
Then the $(2N_f+N_3)$ D-branes which were
going between $(a_1, a_2)$ will recombine with the new $(N_f-N_3)$ D-branes
allowing us to get the total of $3N_f$ D-branes along $[a_1, a_2]$ cycle.
There exist $(3 N_f-N_1-N_2-N_3)$ D-branes wrapping around $[b_3, a_1]$
coming from $(2 N_f-N_1-N_2)$ D-branes between $(b_2, a_1)$ ( that will be
decomposed into the same number of D-branes between $(b_2, b_3)$ and $( b_3, a_1)$ ) 
and $(N_f-N_3)$ D-branes between $(b_3, a_1)$.

The final configuration by putting all together is a configuration of points ordered
as $(b_1, b_2, b_3, a_1, a_2)$ with $(N_f-N_1)$ D-branes wrapped around
$[b_1, b_2]$, $(2 N_f-N_1-N_2)$ D-branes wrapped around $[b_2, b_3]$,
$(3 N_f-\sum_{i=1}^3 N_i)$ D-branes wrapping around $[b_3, a_1]$
and $3 N_f$ D-branes wrapped around $[a_1, a_2]$.
Notice that the number of D-branes along the cycle $[b_3, a_2]$ in the
original configuration of $(a_1, b_1, b_2, b_3, a_2)$ in the moduli space of CY threefolds are the
same of those along the cycle $[a_1, a_2]$ after we moved the points
$b_1, b_2$ and $b_3$.
In the limit 
$b_1 \rightarrow b_3$ and $b_2 \rightarrow b_3$, this is exactly the dual magnetic description
of the original theory. The gauge 
group \cite{Brodie,BS} is $SU(\widetilde{N_c})=SU(3 N_f-N_c)$ 
since $\sum_{i=1}^3 N_i=N_c$.
This is a marginal deformation of $k=3$ duality of \cite{Kutasov} clarifying
the correspondence between $D_3$ and $A_3$ type singularities.
In addition to the $N_f$ flavors of dual quarks $q^i, \widetilde{q_{\widetilde{i}}}$ 
and
adjoint dual superfields $\overline{X}$ and$\overline{Y}$
we have three singlet chiral superfields $M_i$ which interact with
the dual quarks through the superpotential in the magnetic theory.

We expect that for general value of $k$, the above procedure can be
done similarly. 
The final configuration after all the $b_i, b'_j$ and $b''_l$'s are moved
to the left of $a_1$ keeping the order of them  we get 
is a configuration of points ordered as 
$$(b_1, b_2, b_3, b'_1, b'_2, \cdots, b'_{\frac{3(k-1)}{2}},
 b''_1, b''_2, \cdots, b''_{\frac{3(k-1)}{2}}, a_1, a_2)$$
with 
 $(N_f-N_1)$ D-branes wrapped around
$[b_1, b_2]$, $(2 N_f-N_1-N_2)$ D-branes wrapped around $[b_2, b_3]$,
$\cdots$,
$( (3k-1) N_f-\sum_{i=1}^3 N_i-\sum_{j=1}^{\frac{3(k-1)}{2}} N'_j-
\sum_{l=1}^{\frac{3(k-1)}{2}-1} N''_l )$ D-branes around $[b''_{\frac{3(k-1)}{2}-1}, 
b''_{\frac{3(k-1)}{2}}]$,
$(3k N_f-\sum_{i=1}^3 N_i-\sum_{j=1}^{\frac{3(k-1)}{2}} (N'_j+N''_j ))$ 
D-branes wrapped around
$[b''_{\frac{3(k-1)}{2}}, a_1]$ and $3k N_f$ D-branes wrapped around $[a_1, a_2]$.
In the limit of $b_i, b'_j, b''_l \rightarrow b''_{\frac{3(k-1)}{2}}$, 
the gauge group 
$SU(\widetilde{N_c})=
SU(3k N_f-N_c)$ appears. In this case there are also $3k$ singlet fields, $M_{ij} (
i=1, \cdots, k$ and $ j=1, 2, 3 )$ 
coupled to the magnetic quarks.

{\bf 2) $SO(N_c)$ with two symmetric tensors and $N_f$ vectors
($Sp(N_c)$ with two antisymmetric tensors and
$N_f$ flavors) \cite{BS}:}

We discuss supersymmetric Yang-Mills theory with gauge group $SO(N_c)$
where the fields $X$ and $Y$ are in the $\frac{N_c(N_c+1)}{2}-1$
traceless symmetric tensor representation 
of $SO(N_c)$
and $N_f$ fields $Q^i$ are in the $N_c$ dimensional vector representation
of $SO(N_c)$ ($i=1, \cdots, N_f$). The superpotential is
$\mbox{Tr} X^{k+1}+\mbox{Tr} XY^2$ where $k$ is odd. 
Let us study the configuration of points ordered as
$$\left( a_1, a_2, a_3, a'_1, a'_2, \cdots, a'_{k-1},
 a''_1, a''_2, \cdots, a''_{k-1}, 0, a' \right)$$
 where $a'_i$ and $a''_j$ are present for $k \geq 3$.
Suppose we consider
$\frac{N_i}{2} \mp 2$ D-brane charges around the
cycle $[a_i, 0]$,  
$\frac{N'_j}{2} \mp 2$ D-brane charges around the
cycle $[a'_j, 0]$ and $\frac{N''_l}{2} \mp 2$ D-brane charges around 
$[a''_l, 0]$ for each $i, j$ and $l$
such that $\sum_{l=1}^{3} N_i+\sum_{j=1}^{k-1} ( N'_j+ N''_j)=N_c$. 
There are also D-brane charges on the $[0, a']$ cycle of $\frac{N_f}{2}$ for each $i$ and
$\frac{3 N_f}{4}$ for
each $j$ and  $l$ after the action
of the orientifolding on the D-branes.

Let us first analyze the simplest case for the case of $k=1$.
That is the configuration points of ordered as  $( a_1, a_2, a_3, 0, a')$.
We wrap $N_1$ D-branes around $[a_1, 0]$, $N_2$ D-branes around 
$[a_2, 0]$ and $N_3$ D-branes around $[a_3, 0]$.
Each of them has $N_f$ D-branes around $[0, a']$ such that
$\sum_i^3 N_i=N_c$ which can be understood that the number of D-branes on the
cycle $[a_3, 0]$ should be $N_c$ as $a_1$ and $a_2$ get close to $a_3$.
Now the conjugation 
$ (x,y,x', y' z) \mapsto (x^*, y^*, x'^*, y'^*, z^*),$
together with exchange of left- and right-movers on the world sheet,
will produce an orientifolding of the above configuration. The conjugation
preserves the above equation for $a_i$ and $a'$ real. In view of 
type I' theory,
these $D$-branes must be counted as $\frac{N_i}{2}$
and 
$\frac{N_f}{2}$ $D$-branes after orientifolding since the orientifolding
leaves these cycles invariant.
Therefore we get D-brane charges of the $[a_i, 0]$ cycle of $\frac{N_i}{2}\mp 2$.
Here the factor $\mp 2$ is  due to a contribution 
from the orientifold plane in addition to the physical D-branes.
The upper sign corresponds to the $SO(N_c)$ gauge group and
the lower one does the $Sp(N_c)$ gauge group. 

By passing the point $a_3$ through
the point $0$ directly along the real axis due to the fact that
this operation should keep the orientifolding, 
the D-brane charge gets changed to the value of 
$\frac{N_f}{2}-\frac{N_3}{2}\pm 2$ on $[0, a_3]$ where
we assumed $N_f \geq N_3 \mp 4$. Next we move the point
$a_2$ to the positive real axis. 
The D-brane charge changes to the value of 
$-\frac{N_2}{2}\pm 2$ on $[0, a_2]$.
The final configurations after we move $a_1$ are given
by the $(-\frac{N_1}{2}\pm 2)$ D-brane charge on $[0, a_1]$, 
 $(-\frac{N_2}{2}\pm 2)$ D-brane charge on $[0, a_2]$, 
$(\frac{N_f}{2}-\frac{N_3}{2}\pm 2)$ D-brane charge
around $[0, a_3]$, $\frac{N_f}{2}$ D-brane charge on $[a_3, a']$ and
two $\frac{N_f}{2}$ D-brane charges on $[0, a']$. Recombining the last two
$\frac{N_f}{2}$ D-branes into the D-branes around $[0, a_3]$ cycle and taking
the limit of $a_1 \rightarrow a_3$ and $a_2 \rightarrow a_3$, it leads to 
$\frac{3N_f}{2}-\frac{N_1}{2}-\frac{N_2}{2}-\frac{N_3}{2}+6$ 
that shows our expression for the magnetic
dual group \cite{BS} for $k=1$ case,
$SO(\widetilde{N_c})= SO(3 N_f-N_c + 12)$
since there is no orientifold plane for $a_i(i=1, 2, 3)> 0$ and all these
D-brane charges are physical D-branes. There exist also $\frac{3 N_f}{2}$ D-branes
on $[a_3, a']$.
The fields $X$ and  its dual $\overline{X}$ are massive and can be integrated out.
This gives a quartic superpotential for $Y$ and its magnetic dual $\overline{Y}$ which are related
to $A_3$ type singularity appeared in \cite{Intril,ILS}.
For the case of $Sp$ group with two antisymmetric tensors and $N_f$
flavors, by recognizing that in the 
convention of
\cite{ILS} the symplectic group whose fundamental representation is 
$2N_c$ dimensional
as $Sp(N_c)$ and a flavor of it has two fields in the fundamental 
representation therefore $2 N_f$ fields
we obtain $2 \widetilde{N_c}=3(2N_f)-2 N_c -12$ which gives rise to
the following dual description $Sp(\widetilde{N_c})= Sp(3 N_f-N_c - 6)$.
 
For general value of $k$,
the final configurations, after we moved all the $a_i, a'_j$ and $a''_l$'s to the right of
the position of zero by successively doing similar things for the previous
case, are given
by the $(-\frac{N_i}{2}\pm 2)$ D-brane charge on $[0, a_i]$,
the $(-\frac{N'_j}{2}\pm 2)$ D-brane charge on $[0, a'_j]$,
the $(-\frac{N''_l}{2}\pm 2)$ D-brane charge on $[0, a''_l]$ for $1 \leq l \leq k-2$,
$(\frac{3k N_f}{2}-\frac{1}{2}(\sum_{i=1}^3 N_i+
\sum_{j=1}^{k-1}( N'_j+
N''_j) )\pm 2(2k+1))$ D-brane charge
around $[0, a''_{k-1}]$ and $\frac{3k N_f}{2}$ D-brane charge on $[a''_{k-1}, a']$.
Here the factor of $\frac{3k N_f}{2}$ comes from the contributions of
the sum of $\frac{3 N_f}{2}$ and $\frac{3(2k-2) N_f}{4}$.
In the limit of $a_i, a'_j, a''_l \rightarrow a''_{k-1}$,
we get the magnetic dual group, $SO(\widetilde{N_c})=SO(3k N_f-N_c+4(2k+1))$.
By similar reasoning for the counting of dimension and number of fields  of 
$k=1$, we obtain the magnetic dual 
gauge group
for symplectic group as $Sp(\widetilde{N_c})=Sp(3k N_f-N_c-2(2k+1))$.

{\bf 3) $SO(N_c)$ with a symmetric tensor and an adjoint field and $N_f$ vectors
($Sp(N_c)$ with an antisymmetric tensor and 
an adjoint field and $N_f$ vectors) \cite{BS}:}

We analyze supersymmetric Yang-Mills theory with gauge group $SO(N_c)$
where the field $X$ is in the $ \frac{N_c(N_c+1)}{2}-1$ dimensional symmetric
tensor representation of $SO(N_c)$,
the adjoint field $Y$ is in the $\frac{N_c(N_c-1)}{2}$
dimensional antisymmetric tensor  
of $SO(N_c)$ and
$N_f$ flavors $Q^i$ are in the $N_c$ dimensional vector representation
of $SO(N_c)$ ($i=1, \cdots, N_f$). The superpotential is
$\mbox{Tr} X^{k+1}+\mbox{Tr} XY^2$ where $k$ is odd.
Let us consider  the configuration of points ordered as
$$\left( b_1, a_1, a'_1, a'_2, \cdots, a'_{k-1},
 a''_1, a''_2, \cdots, a''_{k-1}, 0, a' \right).$$
After orientifolding,
the net D-brane
charge of $[a_1, 0]$ cycle becomes $\frac{N_0}{2}\mp 2$ and that of $[0, a']$
cycle is $\frac{N_f}{2}$ and we bring other D-branes to the left
hand side of the point $a_1$ in the real axis whose 
D-brane charges of $[b_1, 0]$
cycle are $N_1$ and that of $[0, a']$ cycle are $N_f$.
There are also $(\frac{N'_j}{2} \mp 2)$ D-brane charges around the
cycle $[a'_j, 0]$ and $(\frac{N''_l}{2} \mp 2)$ D-brane charges around 
$[a''_l, 0]$ for each $j$ and $ l$
such that $N_0+2 N_1+\sum_{j=1}^{k-1} (N'_j+N''_j)=N_c$. 
There are also D-brane charges on the $[0, a']$ cycle of 
$\frac{3 N_f}{4}$ for
each $j, l$ after the action
of the orientifolding on the D-branes.

For simplicity, we will start with the case of $k=3$ ( Remember that for the case of $k=1$,
the configuration for the magnetic theory was 
$-N_1$ D-brane charge around the cycle $[0, b_1]$, 
$ \frac{3 N_f}{2}-\frac{N_0}{2} \pm 2$ on $[0, a_1]$
and $\frac{3 N_f}{2}$ on $[a_1, a']$. We have seen that this duality was given in 
\cite{LS,ILS} already ) which is the configuration
of points ordered as
$(b_1, a_1, a'_1, a'_2, a''_1, a''_2, 0, a')$.
We push the point $b_1, a_1, a'_i$ and $a''_j$'s along 
the real axis to the right and pass the point 
$0$.
In order to count the number of D-branes wrapping around cycles
we use the D-brane charge conservation and the orientation of
the D-branes. Then we get 
the $(-\frac{N_1}{2}\pm 2)$ D-brane charge on $[0, a_1]$, 
$\cdots$,  
the $(-\frac{N''_1}{2}\pm 2)$ D-brane charge on $[0, a''_1]$, 
$(\frac{7 N_f}{2}-\frac{1}{2}(2 N_1+
\sum_{j=1}^2 (N'_j+N''_j ))\pm 10)$ D-brane charge
around $[0, a''_2]$ and $\frac{7 N_f}{2}$ D-brane charge on $[a''_2, a']$.
Next we move the point $b_1$ along the real axis from negative to
positive values. The D-brane charge on $[0, b_1]$ is $-N_0$ due to the
orientation.
The $ N_f $ D-branes which were going between $(0, a')$ can be decomposed
into $N_f$ D-branes between $(0, a''_2)$ and 
$ N_f$ D-branes between $(a''_2, a')$.
Therefore the final picture we end up with is that there are
$-N_0$ D brane charge on $(0, b_1)$,  
$(-\frac{N_1}{2}\pm 2)$ D-brane charge on $[0, a_1]$, $\cdots$,
$(\frac{9 N_f}{2}-\frac{1}{2}(N_0+2 N_1+
\sum_{j=1}^2 ( N'_j+N''_j) )\pm 10)$ D-brane charges
on $(0, a''_2)$ and $\frac{9 N_f}{2}$ on $(a''_2, a')$. 
In the limit of $b_1, a_1, a'_i, a''_j \rightarrow a''_2$,
the magnetic dual group can be written as $SO(\widetilde{N_c})=
SO(9  N_f-N_c+20)$. 
By twicing the $N_f$ and $N_c$ and dividing by two which leads to
$\frac{9(2N_f)-2N_c-20}{2}$,
we get $Sp(\widetilde{N_c})=Sp(9 N_f-N_c-10)$ for the symplectic group.

For the general value of $k \geq 5$, the final configuration is that there are
$-N_0$ D brane charges on $(0, b_1)$,
$(-\frac{N_1}{2}\pm 2)$ D-brane charge on $[0, a_1]$, $\cdots$,
$(-\frac{N'_i}{2}\pm 2)$ D-brane charge on $[0, a'_i]$,
$(-\frac{N''_j}{2}\pm 2)$ D-brane charge on $[0, a''_j]$ for $1 \leq j \leq k-2$,
$(\frac{3 k N_f}{2}-\frac{1}{2}(N_0+2 N_1+\sum_{j=1}^{k-1} ( N'_j+N''_j)) \pm 2(2k-1))$
D-brane charges
on $(0, a''_{k-1})$ and $\frac{3 k N_f}{2}$ on $(a''_{k-1}, a')$. 
In the limit of $b_1, a_1, a'_i, a''_j \rightarrow a''_{k-1}$,
the dual theory has gauge group, $SO(\widetilde{N_c})=
SO(3 k N_f-N_c+8k-4)$ and 
$ Sp(\widetilde{N_c})=Sp(3 k N_f-N_c-4k+2)$.

{\bf 4) $SU(N_c)$ with an adjoint field, a symmetric and conjugate symmetric
tensors and $N_f$ fundamental
flavors
($SU(N_c)$ with an adjoint field, an antisymmetric and conjugate antisymmetric
tensors and $N_f$ fundamental flavors)
\cite{BS}:}

The field $X$ is in the adjoint representation of $SU(N_c)$, 
$Y$ and $\widetilde{Y}$ are $ \frac{N_c(N_c+1)}{2}$
symmetric tensor and $\overline{\frac{N_c(N_c+1)}{2}} $ conjugate symmetric tensor
representations of $SU(N_c)$ respectively and there are   
$N_f$ fundamental multiplets $Q^i$ and $N_f$ antifundamental
multiplets $\widetilde{Q}_{\widetilde{i}}$ where $i,
\widetilde{i}=1, \cdots, N_f$.
The superpotential is $\mbox{Tr} X ^{k+1}+\mbox{Tr} XY\widetilde{Y} $ where
$k$ is odd.
Let us consider the configuration of points ordered as
$$\left( b_1, b'_1, b'_2, \cdots, b'_{\frac{3(k-1)}{2}},
 b''_1, b''_2, \cdots, b''_{\frac{3(k-1)}{2}}, a, 0, a' \right).$$
Now we continue to repeat the procedure we have done so far for the
case of ordered configuration as $(b_1, b'_1, b'_2, b'_3, b''_1,
b''_2, b''_3, a, 0, a')$ when we
consider $k=3$ case (Recall that when $k=1$, the configuration
was $-N_1$ D-brane charge on $[0, b_1]$ and $ \frac{3 N_f}{2}-\frac{N_0}{2} 
\pm 2$ on $[0, a]$
and $\frac{3 N_f}{2}$ on $[a, a']$. This duality was discussed in \cite{ILS}). 
After orientifolding,  D-brane charges are $N_1$ around 
$[b_1, 0]$, $\frac{N_0}{2}\mp 2$ around $[a, 0]$, 
$N'_i$ around $[b'_i, 0]$ 
and $N''_j$ around $[b''_j, 0]$.
Each of them produces $\frac{N_f}{2} $ D-brane charge around
 $[0, a']$ except that $N_1$  D-brane on $[0, b_1]$ does $N_f$ D-brane
charge where $N_0+2 N_1+2 \sum_{i=1}^3 ( N'_i + N''_i)=N_c$.
The final configuration is given
by 
$-N_1$ D-brane charge on $(0, b_1), \cdots,$
$(\frac{9 N_f}{2}-\frac{N_0}{2} - N_1-\sum_{i=1}^3 ( N'_i + N''_i)
\pm 2)$ 
D-brane charge on $[0, a]$ and 
$\frac{9 N_f}{2}$ D-brane charge on $[a, a']$.
Finally we see that
the dual theory has the gauge group $SU(\widetilde{N_c})=SU(9 N_f-N_c+4)$.
On the other hand, when we consider antisymmetric and its conjuagte tensors
with flavors 
one
we get $SU(\widetilde{N_c})=SU(9 N_f-N_c-4)$.

For the general value of $k$, after orientifolding 
D-brane charges are $N_1$ around 
$[b_1, 0]$, $\frac{N_0}{2}\mp 2$ around $[a, 0]$, 
$N'_i$ around $[b'_i, 0]$ 
and $N''_j$ around $[b''_j, 0]$.
The final configuration after we move all the $b_1, b'_i, b''_j$ and $a$ to the
positive real axis of $z$-plane is given
by 
$-N_1$ D-brane charge on $(0, b_1), \cdots,$
the $(\frac{3k N_f}{2}-\frac{N_0}{2} - N_1-\sum_{i=1}^{\frac{3(k-1)}{2}} (  N'_i +
N''_i)\pm 2)$ 
D-brane charge on $[0, a]$ and 
$\frac{3 k N_f}{2}$ D-brane charge on $[a, a']$.
We arrive at the magnetic dual group for $SU$,
$SU(\widetilde{N_c})=SU(3 k N_f+4 -N_c)$ and that for $SU(N_c)$ with
an antisymmetric flavor and $N_f$ fundamental flavors is 
$SU(\widetilde{N_c})=SU(3 k N_f-4-N_c)$.

{\bf 5) $SU(N_c)$ with an adjoint field, an antisymmetric tensor and a conjugate 
symmetric tensor \cite{BS}:}

In this case the field $X$ is in the adjoint representation of $SU(N_c)$,
the field $Y$ is in the $\frac{N_c(N_c-1)}{2}$ representation,
the field $\widetilde{Y}$ in the $\overline{\frac{N_c(N_c+1)}{2}}$
representation and $m_f(\widetilde{m_f})$ fields                            
$Q^i(\widetilde{Q_{\widetilde{i}}})$ in the
(anti)fundamental representation. The superpotential is given by
$\mbox{Tr}(X\widetilde{X})^{k+1}+\mbox{Tr} XY\widetilde{Y}$
where $k$ is odd or even.
Let us consider  the configuration of points ordered as
$$\left( a_1, b_1, b'_1, \cdots, b'_{k-1}, a_2 \right).$$
We wrap $N_1$ D-branes around the cycle $[a_1, b_1]$, $N'_i$ D-branes around the 
cycle $[a_1, b'_i]$ for each $i=1, 2, \cdots, k-1$ 
such that $ N_1+\sum _{i=1}^{k-1} N'_i=N_c$,
$\frac{3(m_f+\widetilde{m_f})}{2}$ D-branes around the cycle $[b_1, a_2]$
and $\frac{3(m_f+\widetilde{m_f})}{2}$ D-branes around the cycle $[b'_i, a_2]$ for each $i$.
We expect that the above procedure of case 1)
can be applied similarly, for example, $k=2$. 
We wrap  $N_1$ D-branes around the three cycle
$[a_1, b_1]$ and $N'_1$ D-branes around $[a_1, b'_1]$ such that $N_1+N'_1=N_c$ 
and $\frac{3(m_f+\widetilde{m_f})}{2}$
D-branes around the cycle $[b_1, a_2]$ and $[b'_1, a_2]$ respectively.
The final configuration after we move $b_1$ and $b'_1$ to the left
of $a_1$ in the configuration of point ordered as $(a_1, b_1, b'_1, a_2)$
we get 
is a configuration of points ordered
as $(b_1, b'_1, a_1, a_2)$ with $(\frac{3(m_f+\widetilde{m_f})}{2}- N_1)$ 
D-branes wrapped around
$[b_1, b'_1]$ and $(3 (m_f+\widetilde{m_f})-  N_1-  N_2)$ 
D-branes wrapped around $[b'_1, a_1]$,
$3 (m_f+\widetilde{m_f})$
D-branes wrapped around $[a_1, a_2]$. In the limit 
$b_1 \rightarrow b'_1$, the gauge group 
becomes $SU(\widetilde{N_c})=
SU(3(m_f+\widetilde{m_f})-N_c)$. 

For general value of $k$, 
the final configuration is a configuration of points ordered
as $(b_1, b'_1, \cdots, b'_{k-1}, a_1, a_2)$ with 
$(\frac{3(m_f+\widetilde{m_f})}{2}- N_1)$ D-branes wrapped around
$[b_1, b'_1]$, $(3(m_f+\widetilde{m_f})- N_1- N_2)$ 
D-branes wrapped around $[b'_1, b'_2], \cdots,
 (\frac{3k(m_f+\widetilde{m_f})}{2}-\sum_{i=1}^k N_i)$ D-branes wrapped around
$[b'_{k-1}, a_1]$, and $\frac{3 k(m_f+\widetilde{m_f})}{2}$ 
D-branes wrapped around $[a_1, a_2]$. In the limit 
$b_1, b'_i \rightarrow b'_{k-1}$, the gauge group 
$SU(\widetilde{N_c})=
SU(\frac{3 k(m_f+\widetilde{m_f})}{2}-N_c)$ appears. 

I thank Kyungho Oh for helpful discussions and Dept. of Physics, Hanyang Univ. for the
hospitality where part of this work was done.

\end{document}